\documentstyle[preprint,aps]{revtex}
\input epsf.tex
\def\DESepsf(#1 width #2){\epsfxsize=#2 \epsfbox{#1}}
\newcommand{\eqn}[1]{Eq.~\ref{#1}}
\newcommand{\be}{\begin{equation}}
\newcommand{\ee}{\end{equation}}
\begin{document}
\preprint{\vbox{\hbox{}}} \draft \title{The Topological AC and HMW
Effects, and the Dual Current in
2+1 Dimensions}
\author{
$^1$Xiao-Gang He and $^2$Bruce H. J. McKellar}

\address{
$^1$Department of Physics, National Taiwan University, Taipei,
Taiwan, 10617\\ E-mail: hexg@phys.ntu.edu.tw\\ and\\ $^2$School of
Physics, University of Melbourne, Parkville, Vic 3052, Australia\\
E-mail: b.mckellar@physics.unimelb.edu.au}

\date{September 2000}
\maketitle
\begin{abstract}
We study the Aharonov-Casher (AC) effect and the related He-McKellar-Wilkens
(HMW) effect in $2+1$
dimensions.  In this restricted space these effects are the result of
the interaction of the electromagnetic field tensor with the dual of
a current.  Transferring the dual operation from the current to the
field tensor shows that 
this interaction may be reinterpreted as due to the interaction of an
effective vector potential and a current, and the AC and HMW effects
follow immediately.  A general proof of this for
particles with an arbitrary spin is provided.

The restriction to $2+1$ dimensions, with this interpretation,
provides a unified way of treating the AC and HMW effects for an
arbitrary spin.  Perhaps more interestingly the treatment shows that a
spin-0 particle can show AC and HMW effects, although it has no
magnetic or electric dipole moment in the usual sense.
\end{abstract}


\section{Introduction}
The study of topological phases in quantum mechanics has provided deep
understanding of quantum systems.  A particularly interesting case of
a topological phase is the Aharonov-Bohm (AB) effect, discovered in
1959 by Aharonov and Bohm\cite{1}.  The AB effect is due to a charged
particle traveling in electric and magnetic field free region,
developing a topological phase which influences interference patterns.
Although the electromagnetic field in the region traversed by the
particle vanishes there is a non-zero vector potential in that region
--- in quantum mechanics a vector potential can produce observable
effects.  The AB effect has been observed experimentally\cite{2}.  In
1984 Aharonov and Casher discovered\cite{3} another configuration
where a topological phase can develop, giving rise to what is now
called the Aharonov-Casher (AC) effect.  In this case, the effect
occurs for a neutral spin-1/2 particle with anomalous magnetic dipole
moment interacting with a two dimensional electric field.  This is
again a non-classical effect and a topological phase is developed.
This effect has also been observed experimentally\cite{4}.  In 1993
another topological phase related to electric dipole interaction of a
neutral spin-1/2 particle was discussed by He and MecKellar\cite{7}
and a year later, independently by Wilkens\cite{8}.  This effect, the
He-McKellar-Wilkens (HMW) effect, is the dual of the AC effect with
the replacements of electric field by magnetic field, and electric
charge density by magnetic monopole density.  Since this effect
involves magnetic monopoles, it may be difficult to study it
experimentally.  However, in 1998, Dowling, Williams and
Franson\cite{9} pointed out that the HMW effect can be partially
tested using metastable hydrogen atoms.

All the phases mentioned have some similarities and some of them are
dual to each other\cite{7,8,9,5,6,6a,6b}.  The AC and HMW effects are
however different in many ways to the AB effect.  In particular, the
AC and HMW effects have been shown to be phenomena involving two
spatial dimensions in an essential way\cite{6,11}.  For this reason we
study the AC and HMW effects in 2+1 dimensions directly, in the hope
that this will reveal some of their hidden properties.  In this paper
we report the results of this study.  We show that in 2+1 dimensions
the magnetic and electric dipole interactions for the AC and HMW
effects can be interpreted as due to a dual current interacting with
the electric and magnetic field, respectively.  This way of viewing
the AC and HMW effects provides an easy way to understand the
topological nature of these effects and the dual nature of these two
effects, and to study these effects for other spin systems in an
unified way.  We provide a general proof for arbitrary spin, and give
specific examples using spin-1/2, spin-1 and spin-0 particles.  An
interesting consequence of this new understanding is that even a
spin-0 particle, where no dipole moment can be defined in the ordinary
sense, can have AC and HMW effects.

\section{The AC, HMW Effects and the Dual Current}

We begin by studying the familiar case of the AC
effect of spin-1/2 particles, but restricting
considerations to $2+1$ dimensions from the beginning.  The
Lagrangian for a neutral particle of spin-1/2 with an anomalous
magnetic dipole moment $\mu_m$ interacting  with the
electromagnetic field is given by

\begin{eqnarray}
L = \bar \psi i \gamma^\mu \partial_\mu \psi - m\bar \psi \psi -
{1\over 2} \mu_m\bar \psi \sigma^{\mu\nu} \psi F_{\mu\nu}.
\label{ac}
\end{eqnarray}
The last term in
the Lagrangian is responsible for the AC effect.

We will use the following conventions for the 2+1 dimensional metric
$g_{\mu\nu}$ and the anti-symmetric tensor $\epsilon_{\mu\nu\alpha}$:
\begin{equation}
    g_{\mu\nu} =
\mbox{diag}(1,-1,-1) \quad \quad \mbox{and} \quad \quad
\epsilon_{012} = +1.
\end{equation}
While it is possible to find $2 \times 2$ matrices satisfying the 2+1
dimensional Dirac algebra, there are two inequivalent representations
which generate different Clifford algebras.  In particular they may be
distinguished by the sign in the relationship $\sigma_{12} = \pm
\gamma^{0}$.  If we use the two component formalism we need these two
inequivalent representations of the Dirac matrices to be able to
represent both spin up and spin down particles.  To avoid this
prescription we will use a four compoment Dirac spinor which
can describe spin up and down in the notional $z$ direction for a
particle and for its anti-particle.  In 2+1 dimensions these Dirac
matrices satisfy the following relation:
\begin{equation}
    \gamma^\mu \gamma^\nu = g^{\mu\nu} +
is\epsilon^{\mu\nu\lambda}\gamma_\lambda,
\end{equation}
where $s=-i\gamma^0 \gamma^1 \gamma^2 = -\gamma^0\sigma^{12}$ which
has eigenvalues $\hat s = \pm$.  A particular representation is
\begin{eqnarray}
\gamma^0 &=& \left (\begin{array}{cc}
                 \sigma_3&0\\
                 0&\sigma_3\\
                \end{array}
            \right ),\;\;
\gamma^1 = \left (\begin{array}{cc}
                 i\sigma_2&0\\
                 0&-i\sigma_2\\
                \end{array}
            \right ),\;\;
\gamma^2 = \left (\begin{array}{cc}
                 i\sigma_1&0\\
                 0&i\sigma_1\\
                \end{array}
            \right ),\nonumber\\
\end{eqnarray}
in which the operator $s$  is diagonal and is given by
\begin{eqnarray}
s  &=& \left (\begin{array}{cc}
                 I&0\\
                 0&-I\\
                \end{array}
            \right ).
\end{eqnarray}

The interaction term in the Lagrangian is then
\begin{eqnarray}
 \bar \psi \sigma^{\mu\nu} \psi F_{\mu\nu} =
- F^{\mu\nu} s\epsilon_{\mu\nu\lambda} \bar \psi
\gamma^\lambda \psi,\;\;\;\;\mbox{with}
\;\;\;\; F^{\mu\nu} = \left (
\begin{array}{ccc}
0&-E^1&-E^2\\ E^1&0&-B^3\\E^2&B^3&0 \end{array} \right ),
\label{dual}
\end{eqnarray}
where $E^i$ and $B^i$ are the electric and magnetic fields,
respectively. The indices ``1'' and ``2'' indicate the
coordinates
on the $x-y$ plane along the $x$ and $y$ directions. The index ``3''
indicates that the magnetic field in this configuration is
normal to the $x-y$ plane, in the notional $z$ direction.
The interaction is represented as the electromagnetic field
tensor contracted with the tensor dual to the current
$\bar\psi \gamma^{\mu} \psi$.

We define an ``effective vector potential'' $S_{\mu}$ as the
dual of the field strength tensor
\begin{equation}
    S_\mu = (1/2)\epsilon_{\mu\alpha\beta} F^{\alpha \beta},
\end{equation}
and  write the Lagrangian as
\begin{eqnarray}
L =  \bar \psi i \gamma^\mu \partial_\mu \psi - m\bar \psi \psi +
s\mu_m S_\mu \bar \psi \gamma^\mu \psi.
\end{eqnarray}

In the AC configuration, 
the magnetic field vanishes and $E_{1}, E_{2}$ are constant
in time.  Then
$S_\mu = (0, E_2, -E_1)$. The vector $S_\mu$ satisfies
\begin{equation}
    \epsilon^{0\lambda\mu}\partial_{\lambda}S_{\mu}
    =-\partial^{i}E_{i} = -\lambda_e.
\end{equation}
Here $\lambda_e$ is the surface density of electric charge
in the $(x,y)$ plane.  Making a
transformation $\psi' = \mbox{exp}[-is\mu_m \int^{\vec r} \vec S
\cdot d\vec r] \psi$, one finds in $\lambda_e=0$ region,
\begin{eqnarray}
L = \bar \psi' i \gamma^\mu \partial_\mu \psi' - m\bar \psi'
\psi'. \label{phase1}
\end{eqnarray}
This is a Lagrangian for a free particle.
The phase transformation of \eqn{phase1} is able to convert the
interacting dipole system into a free system.  We will refer to this
phase transformation as the AC transformation. The effective vector potential 
 $S^\mu$ can be viewed as a pure ``gauge'' vector potential in charge free regions. 
The phase, relative to
the configuration without any field, developed in the wave function
when the particle travels along a closed path which encircles a total
linear charge $\Lambda_{e}$ is
\begin{eqnarray}
\theta_{AC} = \hat s\mu_m \oint \vec S \cdot d\vec r =- \hat s\mu_m \int_s
(\bigtriangledown\cdot \vec E) ds = -\hat s\mu_m \Lambda_e.
\end{eqnarray}

It is central to the existence of the AC effect that the magnetic
dipole interaction can be written as a interaction between the
electromagnetic field strength $F_{\mu\nu}$ and the dual of the usual
current $\epsilon^{\mu\nu\lambda} j^\lambda$. From
\eqn{dual},
\begin{eqnarray}
L_{AC} =s{\mu_m\over 2}
F^{\mu\nu}\epsilon_{\mu\nu\lambda}j^\lambda.
\end{eqnarray}
For spin-1/2 Dirac particle $j^\lambda = \bar \psi \gamma^\lambda
\psi$ which is the non-vanishing current\cite{footnote}.
This is the origin
of the AC  topological phase representation of the magnetic dipole
interaction in this 2+1 dimensional configuration.

More generally the interaction $F^{\mu\nu} \epsilon_{\mu\nu\lambda}
j^\lambda$ generates a topological phase regardless of the
specific value of the spin of the particle if the AC conditions
required for the electric field is satisfied.  This can be seen by
studying the change of the action $\Delta S$ of the system due to
$L_{AC}$, for a closed trajectory from time 0 to time T for a point
particle with velocity $\vec v \propto \vec{j}$,
\be
\Delta S \sim \int^T_0F^{\mu\nu}
\epsilon_{\mu\nu\lambda }j^\lambda \sim \int^T_0 (\vec S \cdot \vec
v)dt = \oint \vec S\cdot d\vec r \sim \theta_{AC}
\ee
when the AC conditions
are satisfied.

The electric dipole interaction can also be
interpreted as due to a dual current interaction, demonstrating in a
simple and generic way the topological nature of the
HMW effect.

To see this we first need to turn to the usual $3+1$ dimensional
space, where the HMW effect is due to the interaction term
\be
L_{HMW} =
-i{1\over 2} \mu_e \bar \psi \sigma^{\mu\nu}\gamma_5 \psi F_{\mu\nu}
\ee
in the Lagrangian.  Using the identity $\sigma^{\mu\nu} \gamma_5 =
(i/2)\epsilon^{\mu\nu\alpha\beta} \sigma_{\alpha\beta}$, this becomes
\be
-iF_{\mu\nu} \bar \psi \sigma^{\mu\nu}\gamma_5 \psi = \tilde
F^{\mu\nu}\bar\psi \sigma_{\mu\nu} \psi,
\ee
where $\tilde F^{\mu\nu} =
{1\over 2} \epsilon^{\mu\nu\alpha\beta}F_{\alpha\beta}$ is the $3+1$
dimensional dual of the electromagnetic field tensor.

Now go to  2+1 dimensions, where $L_{HMW}$ can further be written as
\begin{eqnarray}
L_{HMW} = - {1\over 2}s \mu_e \tilde F^{\mu\nu}
\epsilon_{\mu\nu\lambda}\bar \psi \gamma^\lambda \psi,\;\;\;\;
\mbox{with}\;\;\;\;\tilde F^{\mu\nu} = \left (
\begin{array}{ccc}
0&-B^1&-B^2\\ B^1&0&E^3\\B^2&-E^3&0 \end{array} \right )
\end{eqnarray}
As was emphasized previously, the HMW effect is the dual of the AC
effect, it is  the interaction between the dual field
strength $\tilde F^{\mu\nu}$ and the same dual current
$\epsilon_{\mu\nu\lambda} j^\lambda$ which enters the AC effect.

Setting $T^\mu =
(1/2)\epsilon^{\mu\alpha\beta} \tilde F_{\alpha \beta}$, one can
write the Lagrangian as
\begin{eqnarray}
L =  \bar \psi i \gamma^\mu \partial_\mu \psi - m\bar \psi \psi
-s\mu_e T_\mu \bar \psi \gamma^\mu \psi.
\end{eqnarray}

For the HMW configuration, with a vanishing electric field, $T_\mu =
(0, B_2, -B_1)$.  The vector $T_\mu$ satisfies, $\epsilon_{0ij}
\partial^{i}T^{j} = - \partial_{i}B^{i} =-\lambda_m$.  Here
$\lambda_m$ is the line density of the magnetic monopole charge.  The
Lagrangian, expressed in terms of $\psi' = \mbox{exp}[is\mu_e \int^{\vec r}
\vec T \cdot d\vec r] \psi$, is in the form of the Lagrangian of a
free particle in a magnetic monopole free region, just as in  the AC case
discussed earlier.  We will refer to this transformation as HMW
transformation.  $T^\mu$ is a pure ``gauge'' field in magnetic
monopole free regions.  Following previous discussions for the AC
effect one finds that when the particle travels along a closed path
on which $\lambda_m=0$, the wave function develops a non-zero HMW phase given
by, $\theta_{HMW} =\hat s\mu_e \Lambda_m$, where $\Lambda_m$ is the linear
monopole density enclosed by the path.

 The above analysis emphasizes  that the AC and HMW effects
have the same origin. In the AC case the topological term is due
to the interaction of the dual current
$\epsilon^{\mu\nu\lambda} j_\lambda$ with electric field, and in
the HMW case the topological term is due to the interaction of the
same dual current with magnetic field.

Writing the AC and HMW interactions in the dual current
interaction forms, $F^{\mu\nu} \epsilon_{\mu\nu\lambda} j^\lambda$
and $\tilde F^{\mu\nu} \epsilon_{\mu\nu\lambda} j^\lambda$, has
several advantages. It enables one to easily understand the
topological nature and the dual nature of these two effects, and
also provides an unified way to study related topological interactions
for other particles with different spins. One just needs to obtain
the  current $j_\mu$ for the particles concerned.

\section{The AC, HMW effects for arbitrary spins in 2+1 dimensions}

The AC effect with arbitrary spin has been studied in Ref.\cite{6a}
for maximal projection of spin, the highest or lowerest spin
projection on the notional z
direction.  Here we show that for arbitrary spins, one can also
interpret the AC and HMW effects for any projection of the spin as
interaction of a dual current with electric and magnetic field,
respectively.  We emphasis that AC and HMW effects exist as exact
results in 2+1 dimensions in the appropriate field configuration, not
only  for the maximal spin projections, but for all projections.
This  is a more general result than that obtained in Ref.\cite{6a}.  An easy
way of proving our result is to use the Bargmann-Wigner formulation for a
field with arbitrary non-zero spin,

\begin{eqnarray}
L = \bar \psi^{S_m}_{\alpha^{(1)}...\alpha^{(2S)}}
\sum_{n=1}^{2S}[\gamma^{(n)}_\mu
i\partial^\mu - m - {1\over 2} {\mu_m}F^{\mu\nu}\sigma_{\mu\nu}^{(n)}]
\psi^{S_m}_{\alpha^{(1)}...\alpha^{(2S)}},
\label{arb}
\end{eqnarray}
for the AC effect. For the HMW effect one needs to replace $\mu_m F^{\mu\nu}$
by $-\mu_e \tilde F^{\mu\nu}$. In \eqn{arb}
$S$ labels the total spin of the particle, $S_m$ indicates the spin projection
along the notional $z$ direction in the rest frame of the particle.
$\psi^{S_m}_{\alpha^{(1)}...\alpha^{(2S)}}$
is the wave function symmetric under exchange of $\alpha^{(n)}$, and the
$\gamma^{(n)}_\mu$ and $\sigma_{\mu\nu}^{(n)}$ act on the nth component of
the Dirac index of $\psi^{S_m}_{...\alpha^{(n)}...}$.

Using the relation $F^{\mu\nu} \sigma_{\mu\nu}^{(n)} =
- F^{\mu\nu}
s^{(n)} \epsilon_{\mu\nu\lambda}\gamma^{\lambda{(n)}}$,
one can write the Lagrangian as
\begin{eqnarray}
L = \bar \psi^{S_m}_{\alpha^{(1)}...\alpha^{(2S)}}
\sum_{n=1}^{2S}[\gamma^{(n)}_\mu
i\partial^\mu - m I^{(n)} + {\mu_m} S^\mu s^{(n)} \gamma^{(n)}_\mu]
\psi^{S_m}_{\alpha^{(1)}...\alpha^{(2S)}},
\end{eqnarray}
which is very suggestive that the dipole interaction has a form
similar to the dual current form discussed in the previous section.
However it is not quite the same yet, one needs further to show that
after making a AC transformation the Lagrangian expressed in the
transformed field is the same as the Lagrangian for a free particle.
We now show that this is indeed possible.

To this end we note that each of the $s^{(n)}$ commute with the
operator $ \sum_{n=1}^{2S}[\gamma^{(n)}_\mu i\partial^\mu - m I^{(n)}
+\mu_m S^\mu s^{(n)} \gamma^{(n)}_\mu]$.  Therefore if
$\psi^{S_m}_{\alpha^{(1)}...\alpha^{(n)}...}$ is a solution of equation
of motion, $\Sigma \psi^{S_m}_{\alpha^{(1)}...\alpha^{(n)}...}$, where
$\Sigma = \sum_{n=1}^{2S} s^{(n)}$, is also a solution.  However, the
function obtained by the action of an individual $s^{(n)}$ on a
solution of the equation of motion is not generally a solution of the
the equation of motion.  The reason for this is the symmetrisation of
the wave function over the indices, and the exceptional cases are just
those with a maximal spin projection.  For each of the Dirac indices,
$s^{(n)}$ has eigenvalues $\hat s^{(n)} =\pm$, when symmetrize the
wave function, there are $2S+1$ ways to symmetrize the wave function
depending on how many $\hat s^{(n)}$ have ``+'' and ``$-$''.  These
give the wave function for each value of $S_{m} = \sum_{n} \hat
s^{(n)}$.  In particular the eigenfunctions of $\Sigma$ can be constructed 
as 
\begin{eqnarray}
&&\psi^{S_{2S}}_{...\alpha^{(n)}...} \sim \psi^{S_{2S}}_{+...+...+},\nonumber\\
&&\psi^{S_{2S-1}}_{...\alpha^{(n)}...} \sim \psi^{S_{2S-1}}_{-+...+...+}
+ \psi^{S_{2S-1}}_{+-+...+...+} + ...
+ \psi^{S_{2S-1}}_{+...+ ...+-},\nonumber\\
&&...\nonumber\\
&&\psi^{S_{-2S}}_{...\alpha^{(n)}...} \sim \psi^{S_{-2S}}_{-...-...-}.
\end{eqnarray}
Here the subindices $\pm$ indicate the eigenvalues of $s^{(n)}$ on
each individual component.  It is clear that
$\psi^{S_m}_{...\alpha^{(n)}...}$ are not eigenfunction of each
individual $s^{(n)}$, but are eigenfunctions of $\Sigma$ with
eigenvalues $S_m$: $2S$, $2S-1$, ...  $-2S$.  There are total $2S+1$
eigenstates which accounts for all the spin projections.  Because
these properties, one can further rewrite Eq.\ref{arb} as
\begin{eqnarray}
L&& =
\bar \psi^{S_m}_{\alpha^{(1)}...\alpha^{(2S)}}
\sum_{n=1}^{2S}[\gamma^{(n)}_\mu
i\partial^\mu -m I^{(n)}]
\mbox{exp}^{-i(\mu_m/2S)
\sum_{n=1}^{2S} s^{(n)} \int^{\vec r} \vec S\cdot \vec r)}
\psi^{S_m}_{\alpha^{(1)}...\alpha^{(2S)}}\nonumber\\
&&=\bar \psi^{S_m}_{\alpha^{(1)}...\alpha^{(2S)}}
\sum_{n=1}^{2S}[\gamma^{(n)}_\mu
(i\partial^\mu + {\mu_m\over 2S} S^\mu \Sigma)
-m I^{(n)}]
\psi^{S_m}_{\alpha^{(1)}...\alpha^{(2S)}}.
\end{eqnarray}
This is the desired form. Making a AC transformation,
$\psi^{S_m\prime}_{\alpha^{(1)}...\alpha^{(2S)}}
= \mbox{exp}[-i(\mu_m/2S) \Sigma \int^{\vec r} \vec S\cdot d\vec r]
\psi^{S_m}_{\alpha^{(1)}...\alpha^{(2S)}}$,
one obtains
\begin{eqnarray}
L = \bar \psi^{S_m\prime}_{\alpha^{(1)}...\alpha^{(2S)}}
\sum_{n=1}^{2S}[\gamma^{(n)}_\mu
i\partial^\mu - m I^{(n)}
]
\psi^{S_m\prime}_{\alpha^{(1)}...\alpha^{(2S)}}.
\end{eqnarray}
The above is true not only just for the maximal projections with
$S_m = \pm S$, but also for other projections. We have obtained a more
general result than that derived in Ref.\cite{6a}.

Using the fact that $\Sigma \psi^{S_m}_{...\alpha^{(n)}...}
= S_m \psi^{S_m}_{...\alpha^{(n)}...}$,
one easily finds the AC phase to be
\begin{eqnarray}
\theta_{AC} = -\mu_m{S_m\over S} \Lambda_e.
\end{eqnarray}
Replacing $\mu_m F^{\mu\nu}$ by $-\mu_e \tilde F^{\mu\nu}$ one obtains the
case for HMW effect and the topological phase is given by
$\theta_{HMW} = \mu_e(S_m/S) \Lambda_m$.

\section{The AC and HMW effects for spin-1 particles}

The AC effect for particles with spin-1 has been studied in
Ref.\cite{6a} using the Bargmann-Wigner formulation, and in more detail
by Swansson and McKellar\cite{10}, using the Proca and Duffin Kemmer
formalisms for the spin one field.  Here we work directly in 2+1
dimensions, to show how the dual current interacting with electric and
magnetic fields induces AC and HMW effects.  Our results are based on
yet another formulation of the Lagrangian.

The Lagranian for a free spin-1 particle is usually written in the
following form
\begin{eqnarray}
L={1\over 2} G^{*\mu\nu}G_{\mu\nu} -{1\over 2}\left [
G^{*\mu\nu}
(\partial_\mu B_\nu - \partial_\nu B_\mu)
+
(\partial^\mu B^{*\nu} - \partial^\nu B^{*\mu})
G_{\mu\nu}
\right ]
+ m^2 B^\mu B_\nu.
\end{eqnarray}
In 2+1 dimensions, one can write, without lost of generality,
$B_\mu$ and $G_{\mu\nu}$ in terms of two other fields\cite{11},
$\phi_{+\mu}$ and $\phi_{-\mu}$ with
\begin{eqnarray}
&&B_\mu = {1\over \sqrt{2m}}(\phi_{+\mu} + \phi_{-\mu}),\nonumber\\
&&G_{\mu\nu} = \left ( {m\over 2}\right
)^{1/2}\epsilon_{\mu\nu\alpha} (\phi_{+}^\alpha -
\phi_{-}^\alpha),
\end{eqnarray}
and one obtains,
\begin{eqnarray}
L = \left [ m\phi^*_{s'\mu}\phi^\mu_{s'} -
{s'\over 2}
\epsilon_{\mu\nu\alpha}(\phi^{*\alpha}_{s'} \partial^\mu \phi_{s'}^\nu +
\phi^\alpha_{s'} \partial^\mu \phi^{*\nu}_{s'})\right ]. \label{s1}
\end{eqnarray}
Here we adopt a summation convention on $s'$, which, when a repeated
subscript,  is summed over $+$ and $-$.

It is usual to write the non-minimal interactions of a spin-1 particle
with the electromagnetic field, when expressed in terms of the fields
$B^\nu$ and $G^{\mu\nu}$, in terms of the two forms\cite{12},
$i\kappa_m B^*_\mu B_\nu F^{\mu\nu}$ and
$i(\tau_m/m^2)G^*_{\nu\alpha} G^\alpha_{\;\; \mu} F^{\mu\nu}$ which
contribute to both the magnetic dipole and electric quadrupole moments.  The
magnetic dipole moment $\mu_m$\cite{12} is $\mu_m = e
(\kappa_m+\tau_m)/2$, and the electric quadrupole moment $Q_e$ is
$Q_e=-e(\kappa_m-\tau_m)/m^2$.  The AC effect is purely due to the
magnetic dipole moment interaction, and in fact does not appear as an
exact result when the
electric quadrupole interaction is non-zero.  Eliminating the electric
quadrupole moment  implies
that $\kappa_m = \tau_m$.  One then can express the magnetic dipole
interaction in terms of the $\phi_\pm$ fields as
\begin{eqnarray}
&&L_{AC} = i{\kappa_m\over m} F_{\mu\nu}
\phi_{s'}^{*\mu}\phi_{s'}^\nu,\nonumber\\
\end{eqnarray}

For the electric dipole interaction, one just replaces $\kappa_mF^{\mu\nu}$ by
$-\kappa_e\tilde F^{\mu\nu}$, and obtains
\begin{eqnarray}
&&L_{HMW} = -i{\kappa_e\over m}
 \tilde F_{\mu\nu} \phi_{s'}^{*\mu}\phi_{s'}^\nu,\nonumber\\
\end{eqnarray}

In order to show that in the AC and HMW configurations that the
magnetic and electric dipole interactions are topological,
one only needs to show that the
dipole interactions can be written as the interactions between
the dual  current,
and the electric and magnetic fields, respectively.
The  current obtained from Eq. \ref{s1} is given by

\begin{eqnarray}
j_\mu = {s'\over 2} \epsilon_{\mu\nu\alpha}(\phi^{*\alpha}_{s'}
\phi_{s'}^\nu - \phi^\alpha_{s'}\phi^{*\nu}_{s'}).
\end{eqnarray}

With this identification of the  current, it can be easily
seen that the dual current $\epsilon_{\mu\nu\alpha} j^\alpha$ can
be written as
\begin{eqnarray}
\epsilon_{\mu\nu\alpha}j^\alpha =- {s'}(\phi^*_{s'\mu}\phi_{s'\nu}
-\phi^*_{s'\nu}\phi_{s'\mu}).
\end{eqnarray}
One then obtains
\begin{eqnarray}
&&L_{AC}
=ig^\prime_{AC}F^{\mu\nu}\epsilon_{\mu\nu\alpha}j^\alpha,\nonumber\\
&&L_{HMW} =ig^\prime_{HMW}\tilde
F^{\mu\nu}\epsilon_{\mu\nu\alpha}j^\alpha,
\end{eqnarray}
where $g^\prime_{AC,HMW} =\mp s'\kappa_{m,e}/2m$. The phases developed when the
AC and HMW conditions are satisfied are,
$\theta_{AC}(s) = s' (\kappa_m/m)\Lambda_e$ and $\theta_{HMW} =
-s' (\kappa_e/m)\Lambda_m$,
respectively. This completes the
proof that the magnetic and electric dipole interactions in the AC
and HMW configurations are topological. If $\kappa_{m,e}$
is not equal to $\tau_{m,e}$, it is not possible to  make a
AC or a HMW transformation such
that the dipole and quadrupole interactions
are transformed away in charge free regions.

\section{The AC and HMW Effects for Spin-0 particles}

For a particle with non-zero spin it was possible to write the
dipole interaction in term of the dual of the current.  This suggests
adopting an alternative procedure --- start with the interaction of
the electromagnetic field with the dual of the current and see where
that leads.  We demonstrate this procedure by studying particles
with spin-0.  One particularly interesting consequence of this new
approach is that spin-0 scalar particles may also have AC and HMW
effects.  At first sight this looks strange because for a spin-0
particle dipole interaction, in the ordinary sense, can not be
defined.  However when the AC and HMW effects are interpreted as due
to dual current interacts with special configurations of field
strength and the dual field strength, it is very natural to have AC
and HMW effects because the current is well defined for a spin-0
particle which carries an additive quantum number (not necessarily an
electric charge).

For a scalar the free Lagrangian is given by
\begin{eqnarray}
L = (\partial^\mu \phi)^* (\partial_\mu \phi) -m^2 \phi^*\phi.
\end{eqnarray}
The  current obtained from the  above is $i(\phi^* \partial_\mu \phi - \phi
\partial_\mu \phi^*)$. One immediately finds the interaction terms
for the AC and HMW effects to be
\begin{eqnarray}
&&L_{AC}= g_{AC} S^\mu i(\phi^* \partial_\mu \phi - \phi
\partial_\mu \phi^*),\nonumber\\ &&L_{HMW}=
g_{HMW}T^\mu i(\phi^* \partial_\mu \phi - \phi \partial_\mu
\phi^*). \label{st}
\end{eqnarray}

The above Lagrange can NOT be transformed to a free scalar
Lagrangian  by the transformations
$\phi' = \mbox{exp}[-ig_{AC} \int^{\vec r} \vec S \cdot d\vec r]
\phi$ and $\phi' = \mbox{exp}[-ig_{HMW} \int^{\vec r} \vec T \cdot d\vec
r] \phi$ for the AC and HMW cases, respectively. In order to
achieve the transformation,  one needs to add terms $g_{AC}^2 S^\mu S_\mu
\phi^*\phi$ and $g_{HMW}^2 T^\mu T_\mu \phi^*\phi$ to the two
terms in eq.(\ref{st}) respectively. These terms look like the
``seagull'' terms introduced in usual the QED theory of spin-0
particles to ensure  gauge
invariance. Here the $g^2_{AC, HMW}$ terms ensure that
the dipole interactions are pure AC and HMW ``gauge''
interactions, respectively.

\section{Discussions and Conclusions}

In this paper we have shown
that in 2+1 dimensions the electric and magnetic dipole
interactions for the AC and HMW effects can be interpreted as a
dual current interacts with electric and magnetic field, respectively.
This way of viewing the AC and HMW effects allows an easy understanding
of the topological nature of these effects, and it provides a
straight forward proof of the effects as exact results in 2+1
dimensions for particles of arbitrary spin.
We have discussed  specific examples using spin-1/2, spin-1,
and spin-0 particles.

In all the cases studied, the AC and HMW interactions can be
transformed away when the conditions for AC and HMW effects are
satisfied, namely, in regions where there are not electric or magnetic
monopole charges the interactions can be viewed as  pure AC and
HMW ``gauge''
interactions.

We have worked with neutral particles for the AC and HMW effects. In fact
the AC and HMW effects are not limited to neutral particles. They also exist
for charged particles. One can easily obtain the formalisms for the
corresponding effects for charged particles by replacing the
partial derivative operator $\partial_\mu$ by
the covariant derivative $\partial_\mu - ieA_\mu$.
In this case the charged
particles also experience the usual electric interaction. The AC and
HMW transformations will not transform the equations of motion into
free particle equations of motion, but transform the dual current interaction
parts away into the wave function and produce topological phases when the
AC and HMW conditions are satisfied\cite{6}.

An interesting consequence of the new understanding discussed in this paper
is that
even spin-0 particles, where no
dipole moment can be defined in the ordinary sense,
can have AC and HMW effects.
One may
ask if the  effects for spin-0 particles can be realized in physical
situations.
It is not difficult to construct such an interaction theoretically. One
easy way of achieving this is to supersymmetrize the AC and HMW effects.
The super partners of the spin-1/2 particles are spin-0 particles. The
AC and HMW phases for the spin-0 particles are directly related to
those for the spin-1/2 particles.
More detailed studies of supersymmetric AC and HMW effects, and related
problems will be presented elsewhere.  It is not clear to us at this
moment that if in certain composite systems in two spatial dimensions
such interaction can actually exist.  The experimental observation of
these effects remains to be studied.

\noindent {\bf Acknowledgments:}

This work was partially supported by the National Science Council of
R.O.C. under grant number  NSC 89-2112-M-002-016 and by the Australian
Research Council under a University of Melbourne Small Grant.  BHJMcK
thanks the Department of Physics at the National Taiwan University for
their hospitality which enabled us to commence this work.
XGH thanks the School of Physics at the University of Melbourne for their 
hospitality
while this work was finalized.


\vspace{1cm}


\begin{thebibliography}{99}

\bibitem{1} Y. Aharonov and D. Bohm, Phys. Rev. {\bf 115}, 485(1959).

\bibitem{2} For a review see, M. Peshkin and A. Tonomura,
The Aharonov-Bohm Effect (Springer-Verlag, Berlin, 1989).

\bibitem{3} Y. Aharonov and A. Casher, Phys. Rev. Lett. {\bf 53}, 319(1984);
See also J. Anandan, Phys. Rev. Lett. {\bf 48}, 1660(1982);
Phys. Lett. {\bf B138}, 347(1989).

\bibitem{4} A. Cimmino et al., Phys. Rev. Lett. {\bf 63}, 380(1989);
K. Sangster et al., Phys. Rev. Lett. {\bf 71}, 3641(1993).

\bibitem{7} X.-G. He and B. H. J. McKellar, Phys. Rev. {\bf A47}, 3424(1993).

\bibitem{8} M. Wilkens, Phys. Rev. Lett. {\bf 72}, 5(1994).

\bibitem{9} J. Dowling, C. Williams and J. Franson, Phys. Rev. Lett.
{\bf 83}, 2486(1999).

\bibitem{5}
T. Boyer, Phys. Rev. {\bf A36}, 5083(1987);
Y. Aharonov, P. Pearle and L. Vaidman, Phys. Rev. {\bf A37},
4052(1988);
A. Goldhaber, Phys. Rev. Lett. {\bf 62}, 482(1989);
C. R. Hagen, Phys. Rev. Lett. {\bf 64}, 2347(1990).

\bibitem{6}
X.-G. He and B. H. J. McKellar, Phys. Lett. {\bf B256}, 250(1991);
ibid {\bf B264}, 129(1991); X.-G. He, in Proceedings of `` Test
of Fundamental Laws in Physics'', Les Arcs, Saviors, France,
Jan 26 - Feb. 2, 1991, p352, Ed. O. Fackler et al., Frontieres.

\bibitem{6a} Y. Azimov and R. Ryndin, JETP Lett. {\bf 61}, 345(1995).

\bibitem{6b} G. Spavieri, Phys.  Rev.  Lett.  {\bf 81}, 1533 (1998);
M. Wilkens, Phys.  Rev.  Lett.  {\bf 81}, 1534 (1998);    H. Liu, X. Y.
Huang, and S. W. Qian, Chinese Phys.  Lett.  {\bf 12}, 327 (1995); J. Yi, G.
S. Jeon, and M. Y. Choi, Phys.  Rev.  {\bf B 52}, 7838 (1995); H. Q. Wei, R.
S. Han, and X. Q. Wei, Phys.  Rev.  Lett.  {\bf 75}, 2071 (1995); C. C.
Chen, Phys.  Rev. {\bf A 51}, 2611 (1995); U. Leonhart and M. Wilkens,
Europhys.  Lett. {\bf 42}, 365 (1998);    C. R. Hagen, Phys.  Rev.
Lett. {\bf 77}, 1656 (1996); H. Q. Wei, X. Q. Wei, R. S. Han, Phys.  Rev.
Lett. {\bf 77}, 1657 (1996).

\bibitem{footnote} Note that a spin-1/2
Majorana particle which carries no quantum numbers has a vanishing
vector current, and vanishing electric and magnetic moments --- this
analysis does not apply to Majorana particles.

\bibitem{10}
C. Hagen and S. Ramaswamy, Phys. Rev. {\bf D42}, 3524(1990).

\bibitem{11} B. H. J. McKellar and J. A. Swansson, preprint
UM-P-032/2000, quant-ph/0007118.

\bibitem{12} K. Kim and Y.-S. Tsai, Phys. Rev. {\bf D7}, 3710(1978).

\bibitem{13} X.-G. He and B. H.J. McKellar, in preparation.

\end{thebibliography}
\end{document}